\documentclass[onecolumn]{mn2e}
\usepackage{graphicx}
\usepackage{amssymb}
\newcommand{\sr}{{\rm sr}}
\newcommand{\cals}{\zeta_s}
\newcommand{\calr}{\zeta_r}

\newcommand{\apjl}{ApJ Lett.}
\newcommand{\mnras}{MNRAS}
\newcommand{\aap}{A\&A}
\newcommand{\aj}{AJ}
\newcommand{\physrep}{Phys.~Rep.}
\newcommand{\araa}{ARAA}
\newcommand{\apj}{ApJ}
\newcommand{\apjs}{ApJ Suppl.}

\title[{Dark energy questions \& weak lensing tomography}]{
Probing decisive answers to dark energy questions from cosmic complementarity
and lensing tomography 
}

\author[{M. Ishak}]{Mustapha Ishak\\Department of Astrophysical Sciences, Princeton University, Princeton, NJ 08544, USA\\mishak@princeton.edu }

\date{\today}

\begin{document}

\maketitle

\begin{abstract}
We study future constraints on dark energy parameters determined 
from several combinations of cosmic microwave background experiments,
supernova data, and cosmic shear surveys with and without tomography.
In this analysis, we look in particular for combinations of experiments
that will bring  the uncertainties to a level of precision tight
enough (a few percent) to answer decisively some of the dark energy questions. In view
of the parameterization dependence problems, we probe the dark energy
using two variants of its equation of state $w(z)$, and its energy
density $\rho_{de}(z)$. For the latter, we model $\rho_{de}(z)$ as a
continuous function interpolated using dimensionless parameters
$\mathcal{E}_i(z_i)\equiv \rho_{de}(z_i)/\rho_{de}(0)$.
 We consider a large set of 13 cosmological
and systematic parameters, and assume reasonable priors on the lensing
and supernova systematics. For CMB, we consider future constraints
from 8 years of data from WMAP,  one year of data from Planck,  and
one year of data from the Atacama Cosmology Telescope (ACT). We use
two sets of 2000 supernovae with $z_{max}=0.8$ and 1.5 respectively,
and consider various cosmic shear reference surveys: a wide
ground-based like survey, covering 70\% of the sky, and with successively
2 and 5 tomographic bins; a deep space-based like 
survey with 10 tomographic bins and various sky coverages. 
The one sigma constraints found are 
$\{\sigma(w_0)=0.086,\sigma(w_1)=0.069\}$,
$\{\sigma(w_0)=0.088,\sigma(w_a)=0.11\}$, and
$\{\sigma(\mathcal{E}_1)=0.029,\sigma(\mathcal{E}_2)=0.065\}$ 
from Planck, supernovae and the ground-based like lensing survey with 2 bins.
When 5 bins are used within the same combination the constraints reduce to 
$\{\sigma(w_0)=0.04,\sigma(w_1)=0.034\}$,
$\{\sigma(w_0)=0.041,\sigma(w_a)=0.056\}$, and
$\{\sigma(\mathcal{E}_1)=0.012,\sigma(\mathcal{E}_2)=0.049\}$.
Finally, when the 
deep lensing survey with 10\% coverage of the sky and 10 tomographic bins 
is used along with Planck and the deep 
supernovae survey, the constraints reduce to
$\{\sigma(w_0)=0.032,\sigma(w_1)=0.027\}$,
$\{\sigma(w_0)=0.033,\sigma(w_a)=0.04\}$, and
$\{\sigma(\mathcal{E}_1)=0.01,\sigma(\mathcal{E}_2)=0.04\}$.
Other coverages of the sky and other combinations of experiments are
explored as well.
Although some worries remain about other systematics, our study
shows that after the combination of the three probes, lensing
tomography with many redshift bins and large coverages of the sky
has the potential to add key improvements to the dark energy parameter constraints.
However, the requirement for very ambitious and sophisticated surveys in 
order to achieve some of these constraints or to improve them 
suggests the need for new tests to probe the nature of dark energy 
in addition to constraining its equation of state.
\end{abstract}
\section{Introduction}
\label{sec:intro}
One of the most important and challenging questions in cosmology and
particle physics is to understand the nature of the dark energy that
is driving the observed cosmic acceleration, see e.g. 
\cite{weinberg,carroll1,turner,sahni,padmanabhan,ishak2005b}.
An important approach to this problem is to constrain 
the properties  of dark energy using cosmological probes. 
This would provide measurements that would allow
one to test various  competing models of dark energy. However, due to
a high degeneracy within a narrow range of the parameter space,
constraining conclusively dynamical dark energy models is going to be
a difficult goal to achieve, and much effort and strategy will be
needed. Ultimately, a combination of powerful cosmological probes and
tests will be necessary. 

In this paper, we study how dark energy parameters are constrained
from different combinations of cosmic microwave background (CMB)
experiments, supernovae of type Ia (SNe Ia) data, and weak lensing
surveys (WL) with and without tomography.  In particular, we look for
combinations of experiments that will  be able to constrain these
parameters well enough to settle  decisively some of the dark energy
questions, say to a few percent. When  CMB measurements constrained the
total energy density to $\Omega_T=1.02\pm 0.02$ to a one sigma level
\cite{Spergel,Bennett}, it became generally more accepted that spatial
curvature is negligible. Thus, an uncertainty of a few percent on dark
energy parameters could be set as a reasonable goal.
Of course, one should bear in mind that it will remain always possible 
to construct dynamical dark energy models that are indistinguishable 
from a cosmological constant within these limits, and therefore, 
one needs to resort to cosmological tests 
beyond the equation of state measurements.
A better scenario providing a decisive answer would be one in which 
one could show that dark energy is clearly not a cosmological constant.

It is certainly wise to probe the nature of dark energy
using gradual steps. 
However, in both the scenarios mentioned above, one should
keep in mind that the results and conclusions obtained from 
an analysis where the equation of state is assumed constant 
are subject to changes if the equation of state is allowed 
to vary with the redshift. 
In this paper, we consider dark energy with a varying equation of state. 

We chose the combination CMB+SN Ia+WL  as various studies have already 
shown that supernovae type Ia  constitute a powerful probe of dark
energy via distance-redshift measurements, see for example 
\cite{Riess98,Garnavich98,Filippenko,Perlmutter98,Perlmutter97,Riess00,Riess01}\\
\cite{Tonry,Knop,Barris,Riess04}. 
Also, several parameter forecast studies have shown that combining
constraints from weak gravitational lensing with constraints from the
CMB  is a powerful combination to constrain dark energy; see, e.g. 
\cite{Hu01,Huterer01,HutererTurner,Benabed}\\
\cite{Abazajian,Refregier,Heavens,Simon,JainTaylor}\\
\cite{BernsteinJain,SongKnox}.
Importantly, weak lensing measurements are sensitive to both the effect of dark
energy on the expansion history and its effect on the growth factor of
large-scale structure. Furthermore, in addition to tightening the constraints,
using independent probes will allow one to test the systematic errors
of each probe, which are serious limiting factors in these
studies. For each of these probes, much data will be available in the
near and far future. For WL, there are many ongoing, planned and
proposed surveys,  such as the Deep Lens Survey ({http://dls.bell-labs.com/})
\cite{2002SPIE.4836...73W}; the NOAO Deep Survey ({\slshape
  http://www.noao.edu/noao/noaodeep/}); the Canada-France-Hawaii
Telescope (CFHT) Legacy Survey ({\slshape http://www.cfht.hawaii.edu/Science/CFHLS/}) \cite{2001misk.conf..540M}; the Panoramic Survey
Telescope and Rapid Response System ({\slshape
 http://pan-starrs.ifa.hawaii.edu/}); 
the {\slshape Supernova Acceleration Probe} ({\slshape SNAP}; {\slshape http://snap.lbl.gov/})
\cite{2003astro.ph..4417R,2003astro.ph..4418M,2003astro.ph..4419R};
and the Large Synoptic Survey Telescope (LSST; {\slshape
  http://www.lsst.org/lsst\_home.html}) \cite{2002SPIE.4836...10T}.
Similarly, there are many ongoing, planned and proposed SNe Ia surveys,
such as the Supernova Legacy Survey \cite{SNLS1,SNLS2} (SNLS);
 The Nearby Supernova Factory (SNfactory) \cite{SNfactory};
the ESSENCE project \cite{ESSENCE1,ESSENCE2,ESSENCE3};
Sloan Digital Sky Survey (SDSS) \cite{SDSS};
The Carnegie Supernova Project \\
\cite{Freedman}; 
and the Dark Energy Camera Project \cite{DECamera1}. We should mention that 
there are other noteworthy cosmological tools for probing dark energy
that we did not consider in this study, notably 
clusters of galaxies (see for example \cite{mohr,wang2004} and references therein), 
Lyman-alpha forests (see for example \cite{mandelbaum,Seljak2004c}), and baryonic oscillations (see for example \cite{Eisenstein2003,Seo2003,Blake2003,Linder2003c}). 

For the CMB, we consider future constraints from  8 years of data from
the Wilkinson Microwave Anisotropy Probe (WMAP-8)
\cite{Bennett,Spergel}, 1 year of data from the Planck satellite (PLANCK1), and
1 year of data from the Atacama Cosmology Telescope (ACT), see
e.g. \cite{Kosowsky}. We use two sets of 2000 supernovae with
$z_{max}$=0.8 and  $z_{max}$=1.5 respectively, and consider two types
of cosmic shear surveys: a ground-based like survey covering 70\% of
the sky with source galaxy redshift distribution  having a median
redshift $z_{med}\approx 1$, and a space-based like deep survey
covering successively 1\%, 10\% and 70\% of the sky with $z_{med} \approx 1.5$. 

We take into account in our analysis systematic 
limits for the supernovae by adding a systematic uncertainty floor
in quadrature following \cite{Kim2003}. We also include for weak 
lensing the effect of the redshift bias and the shear 
calibration bias by adding and marginalizing over the corresponding 
parameters as in \cite{Ishak2004}.

The constraints on the dark energy equation of state are
parameterization dependent; see,
e.g. \cite{wangandtegmark,upadhye}. Also, there is a smearing effect
due to double integration involved when using the equation of state
\cite{maor1,maor2}. In order to partly avoid this, one could probe
directly the variations in the dark energy density using the data;
see, e.g. \cite{wangandpia,wangandfreese}. However, it has been argued
that the equation of state is closer to the physics, as it also
contains information on the pressure, and, trying to probe the
equation of state from the density leads to instability and bias
\cite{linder2004}. Therefore, we choose in this analysis to use both
and parameterize the dark energy using its density $\rho_{de}(z)$ 
as well as two different parameterizations  of its equation of state $w(z)$. 
\section{Cosmological model and dark energy parameterization}\label{sec:model_param}

\subsection{Model}\label{sec:model}
A total set of 13 parameters is considered as follows. For constraints
from WL, we use: $\Omega_{m}h^2$, the physical matter density;
$\Omega_\Lambda$,$w_0$ and $w_1$ (or $w_a$), respectively the fraction
of the critical density in a dark energy component and its equation of
state parameters (see Sec. \ref{sec:de_param} (alternatively, we use
the dark energy density  parameters  $\mathcal{E}_i\equiv
\rho_{de}(z_i)/\rho_{de}(0)$ with $i=1,2$ (see section
\ref{sec:de_param})); we use $n_s(k_0=0.05h/{\rm Mpc})$ and $\alpha_s$,
the spectral index and running of the primordial scalar power spectrum
at $k_0$; $\sigma_8^{\rm lin}$, the amplitude of linear
fluctuations. In order to parameterize some systematics, we include as
a parameter $z_p$, the characteristic redshift of source galaxies (see
Eq. \ref{eq:z_dist}), as well as $\cals$ and $\calr$, the calibration
parameters as defined in Ref.~\cite{Ishak2004} which determine the
absolute calibration error on the lensing power spectrum,
\cite{2003MNRAS.343..459H}), and the relative calibration between
tomography bins (see Sec. \ref{sec:systematics}). When we combine this
with  the CMB, we include $\Omega_{b}h^2$, the physical baryon
density; $\tau$, the optical depth to reionization; $T/S$ the
scalar-tensor fluctuation ratio. We assume a spatially flat Universe
with $\Omega_{m}+\Omega_{\Lambda}=1$. This fixes $\Omega_m$ and $H_0$
as functions of the basic parameters. We do not include massive
neutrinos, or primordial isocurvature perturbations. For the supernova
analysis, we use $\Omega_{\Lambda}$, $w_0$, $w_1$, (or
$\mathcal{E}_1$, $\mathcal{E}_2$) and treat the magnitude parameter
$\mathcal{M}$ as a nuisance parameter. We use the fiducial model
(e.g. Ref. \cite{Spergel} and add $w_0$ and $w_1$): $\Omega_b
h^2=0.0224$, $\Omega_m h^2=0.135$, $\Omega_{\Lambda}=0.73$,
$w_0=-1.0$, $w_1=0.0$, $n_s=0.93$, $\alpha_s=-0.031$,
$\sigma_{8}=0.84$, $\tau=0.17$, $T/S=0.2$, $z_p=0.76,\,1.12$,
$\cals=0.0$, and $\calr=0.0$. 

\subsection{Dark energy parameterization}\label{sec:de_param}

As mentioned earlier, constraining the dark energy using its 
equation of state is known to be parameterization dependent,
e.g. \cite{wangandtegmark,upadhye}, and also suffers from a smearing 
due to the double integration involved \cite{maor1,maor2}. 
Alternatively, one can probe directly the variations of the 
dark energy density as a function of redshift $\rho_{de}(z)$. 
On the other hand, it has been argued that the equation of
state contains information on both the density and pressure 
of the dark energy and using the density to probe the equation of 
state may lead to instability and bias \cite{linder2004}.
We chose to study the constraints on dark energy using both approaches.

\subsubsection{The equation of state}

There are several parameterizations of the dark energy equation of state
that have been used to study currently available data or to do
parameter constraint projections. Discussions of the advantages and
drawbacks of some of these parameterizations can be found in
\cite{wangandtegmark,upadhye}.  We used here the following two
parameterizations, which have no divergence at very large redshift:
\begin{center}
a) {$(w_0,\,w_1)$}
\end{center}
Here $w_1$ represents the redshift derivative of $w(z)$ in the recent
past as follows (see, e.g. \cite{upadhye})
\begin{eqnarray}
\label{eqn:param1}
w(z)         &  = & \left\{ \begin{array}{cl}
w_0 + w_1 z      &       \textrm{if $z<1$} \\
w_0 + w_1        & \textrm{if $z \geq 1$}.
\end{array} \right.
\end{eqnarray}
The evolution of dark energy density with redshift is
given by $\rho_{de}(z) = \rho_{de}(0) \mathcal{E}(z)$
where
\begin{equation}
\label{eq:E_z}
\mathcal{E}(z)  \equiv  \left\{ \begin{array}{cl}
(1+z)^{3(1+w_0-w_1)} \; e^{3w_1z}      &       \textrm{if $z<1$,} \\
(1+z)^{3(1+w_0+w_1)} \; e^{3w_1(1-2 \ln 2)} & \textrm{if $z \geq 1$}.
\end{array} \right.
\end{equation}
\begin{center}
b) {$(w_0,w_a)$}
\end{center}
Here the equation of state is parameterized as  \cite{Chevalier,linder2003a}
\begin{equation}
w(a)=w_0+w_a\frac{z}{1+z}=w_0+w_a(1-a)
\end{equation}
where $a$ is the scale factor. The dark energy density
evolves with $\mathcal{E}(a)$ now given by
\begin{equation}
\mathcal{E}(a)=a^{-3(1+w_0+w_a)}e^{-3w_a(1-a)}.
\label{eq:Ewa}
\end{equation}

\subsubsection{The density parameters: 
 $\mathcal{E}_1\equiv \frac{\rho_{de}(z_1)}{\rho_{de}(0)}$,
 $\mathcal{E}_2\equiv \frac{\rho_{de}(z_2)}{\rho_{de}(0)}$}

Following \cite{wangandpia,wangandfreese}, we parameterize
$\mathcal{E}(z)\equiv \frac{\rho_{de}(z)}{\rho_{de}(0)}$ as a
continuous function interpolated between today and its amplitude
parameters $\mathcal{E}_1$ and $\mathcal{E}_2$ corresponding
respectively to  z=0.5 and 1.0, and remaining constant at higher
redshifts. We use a polynomial interpolation as in
\cite{wangandpia,wangandfreese} so  
\begin{equation}
\mathcal{E}(z)=1+(4\mathcal{E}_1-\mathcal{E}_2-3)\frac{z}{z_{max}}+
2(\mathcal{E}_2-2 \mathcal{E}_1 +1 ) \frac{z^2}{z_{max}^2}.
\end{equation}
where the parameters $\mathcal{E}_1$, and $\mathcal{E}_2$ will be
constrained from the data. As suggested in \cite{wangandpia} and \\
\cite{wangandfreese}, we could use more density parameters
than two as much more data will be available in the future, but we
chose to use only two parameters in order to keep the number of
parameters equal to the equation of state case and be able to make a
fair comparison of the results. Departures of the density parameters
from unity will be an indication of a redshift evolution of the dark
energy density and will rule out a cosmological constant. 
\begin{table}
{
  \begin{center}
  \caption{CMB experiment specifications for Planck and ACT. The parameters used for WMAP 8 years are 
based on the projection of the one year operation and are described in
section 5.}
  \label{tab:cmbexp}
%\begin{ruledtabular}
  \begin{tabular}{ccccccccccc}
\hline
            &  $f_{sky}$  &$l_{max}$& f(GHz)& $\theta_{b}$(arcmin)&$\frac{\Delta T}{T}(10^{-6})$ & $\frac{\Delta P}{T}(10^{-6})$   \\\hline
Planck-1    &  0.8        & 2500    & 100   & 9.5       & 2.5         & 4.0            \\
            &             &         & 143   & 7.1       & 2.2         & 4.2            \\
            &             &         & 217   & 5.0       & 4.8         & 9.8            \\\hline
            &  $f_{sky}$  &$l_{max}$& f(GHz)& $\theta_{b}$(arcmin)&$w_{{\rm eff.}}(\sr^{-1})$         &                           \\\hline
ACT-1       &  0.005      & 8000    & 150   & 1.7               & $3 \times 10^{18}$ &  \\
\hline
  \end{tabular}
%\end{ruledtabular}
 \end{center}
}
\end{table}
\section{Probing dark energy with cosmic shear}\label{sec:wl}
\begin{figure}
\begin{center}
\includegraphics[width=2.3in,angle=-90]{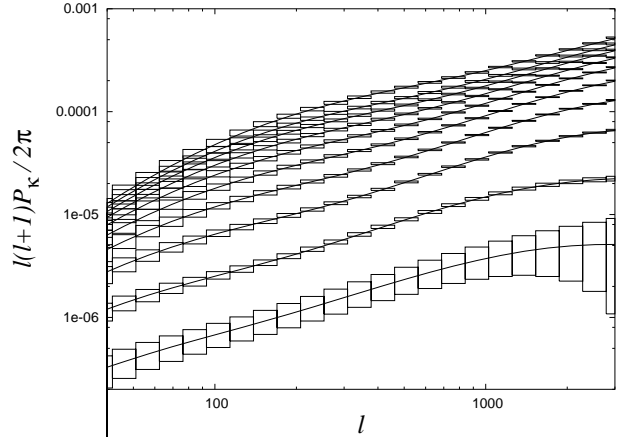}
\caption{\label{fig:convergence} Convergence auto power
  spectra for the 10 tomography bins. All the parameters are fixed at
  their fiducial values. From the bottom to the top, the curves
  correspond to the redshift intervals
  $[0.0,0.3],...,[2.7,3.0]$. For each curve, we display the sample
  variance errors averaged over bands in $l$. 
  }
\end{center}
\end{figure}
Weak lensing is a very promising tool for an era of 
precision cosmology. Already, several studies used currently available
cosmic shear data to constrain various cosmological parameters 
\cite{2003PhRvL..90v1303C,2002A&A...393..369V,2003PhRvD..68l3001W,2003AJ....125.1014J,2004astro.ph..4195M}.
Using statistical inference theory, many other studies showed the
promise of this probe 
 \cite{1999ApJ...514L..65H,Hu01,Huterer01,Abazajian} \\
\cite{Benabed,2004MNRAS.348..897T,2003astro.ph.11104T,Heavens,JainTaylor}\\
\cite{BernsteinJain,Ishak2004,Simon}.
In particular, weak lensing was shown to constrain significantly the
dark energy parameters.  The advantage of weak lensing is that it is
sensitive to the effect of dark energy on the expansion history and
its effect on the growth factor of large-scale structure. Another
advantage of weak lensing is that it allows one to construct new tests
or techniques to probe cosmology. These include redshift bin
tomography \cite{Hu99,Hu02},  cross-correlation cosmography
\cite{BernsteinJain}, and the use of higher order statistics such as
the bispectrum; see, e.g. \cite{2004MNRAS.348..897T}. We explore in
this analysis the constraints obtained from the convergence power
spectrum and multiple bin tomography. 
\subsection{Convergence power spectrum}
\begin{figure}
\begin{center}
\includegraphics[width=2.3in,angle=-90]{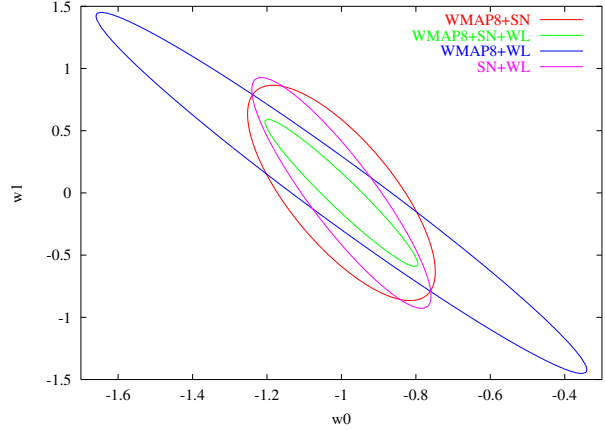}
  \caption{
The $1\sigma$ confidence two-dimensional regions ($\Delta \chi^2=1$)
for the $(w_0,w_1)$ parameters. The plots contrast the
constraints obtained from different combinations of 8 years of data from WMAP,
2000 SNe Ia with $z_{max}=0.8$, and a ground-based like WL reference survey with
$f_{sky}=0.7$, $\bar n = 30\, {\rm gal/arcmin}^2$, 
$\left<\gamma_{int}^2\right>^{1/2} \approx 0.4$, and a median
redshift $z_{med}\approx 1$.
    }\label{fig:wmap}
\end{center}
\end{figure}
Light rays traveling to us from background galaxies get deflected by
mass fluctuations in large scale structures. This results in
distortions of the  sizes and shapes of these galaxies that can be
described by the transformation matrix 
\begin{eqnarray}
A_{ij} \equiv {\partial \theta_s^i \over \partial \theta^j}=
\left(
\begin{array}{cc}
1-\kappa-\gamma_1 & \gamma_2 \\
\gamma_2 & 1-\kappa+\gamma_1
\end{array}
\right)\,\,
\end{eqnarray}
where {\boldmath $\theta_s$} is the angular position in the source plane;
 {\boldmath $\theta$} is the angular position in the image plane;
$\kappa$ is the convergence and describes the magnification
of the size of the image; $\gamma_1$ and $\gamma_2$ are the components
 of the complex shear and describe the distortion of the shape of the
 image. In the weak gravitational lensing limit, $|\kappa|$,
 $|\gamma|$ $\ll 1$.

The convergence is given by a weighted projection of the matter energy
density fluctuations $\delta \equiv \delta \rho/\rho$ along the line
of sight between the source and the observer,
\begin{equation}
\label{eq:kappa}
\kappa(\hat{\theta})=\int^{x_{_H}}_{0}W(\chi)\delta(\chi, \chi
      {\hat{\theta}})d{\chi}
\end{equation}
where $\chi$ is the radial comoving coordinate and $\chi_{_H}$
is the comoving coordinate at the horizon.

The convergence scalar field can be decomposed into multipole moments
of the spherical harmonics as
\begin{equation}
\kappa(\hat{\theta})=\sum_{lm}\kappa_{lm}Y^m_l(\hat{\theta}),
\end{equation}
where
\begin{equation}
\kappa_{lm}=\int d\hat{\theta} \kappa(\hat{\theta},\chi) Y^{m*}_l(\hat{\theta}).
\end{equation}
The convergence power spectrum $P^{\kappa}_{l}$ is then defined by
\begin{equation}
\left < \kappa_{lm} \kappa_{l'm'} \right >=\delta_{ll'} \delta_{mm'} P^{\kappa}_l
\end{equation}
and we will use it as our weak lensing statistic.
In the Limber approximation, it is given by
\cite{1992ApJ...388..272K,1997ApJ...484..560J,1998ApJ...498...26K}:
\begin{equation}
P^{\kappa}_l =
\frac{9}{4} H_0^4\Omega_m^2\int^{\chi_H}_{0}
\frac{g^2(\chi)}{a^2(\chi)}P_{3D}
\left(\frac{l}{\sin_{K}(\chi)},\chi\right) d\chi.
\end{equation}
where $P_{3D}$ is the $3D$ nonlinear power spectrum of the matter
density fluctuation, $\delta$;  $a(\chi)$ is the scale factor; and
$\sin_{K}\chi=K^{-1/2}\sin(K^{1/2}\chi)$ is the comoving angular
diameter distance to $\chi$ (for the spatially flat universe used in
this analysis, this reduces to $\chi$). The weighting function
$g(\chi)$ is the source-averaged distance ratio given by
\begin{equation}
\label{eq:weighting}
g(\chi) = \int_\chi^{\chi_H} n(\chi') {\sin_K(\chi'-\chi)\over
\sin_K(\chi')} d\chi',
\end{equation}
where $n(\chi(z))$ is the source redshift distribution normalized by
$\int dz\; n(z)=1$. For $n(z)$, we use the distribution
\cite{2000Natur.405..143W}:
\begin{equation}
n(z)=\frac{z^2}{2 z_0^3}\, e^{-z/z_0},
\label{eq:z_dist}
\end{equation}
which peaks at $z_p=2z_0$.
For cosmic shear calculations, 
we integrate numerically the growth factor using \cite{linder2003}
\begin{equation}
G''+\left[\frac{7}{2}-\frac{3}{2}\frac{w(a)}{1+X(a)}\right]\frac{G'}{a}+\frac{3}{2}\frac{1-w(a)}{1+X(a)}\frac{G}{a^2}=0 
\end{equation}
where $G=D/a$ is the normalized growth factor with $D=\delta(a)/\delta(a_i)$, 
\begin{equation}
X(a)=\frac{\Omega_m}{(1-\Omega_m)a^3 \mathcal{E}(a)},
\end{equation}
and $\mathcal{E}(a)$ is as given in Eq. \ref{eq:Ewa}.
We use the mapping procedure {\sc halofit}
\cite{smith2003} 
to calculate the non-linear power spectrum. We show in
Fig. \ref{fig:convergence} convergence power spectra for the 10
tomographic bins. We also show the sample variance errors averaged
over bands in $l$. 
\subsection{Weak lensing tomography}
\begin{figure}
\begin{center}
\includegraphics[width=2.3in,angle=-90]{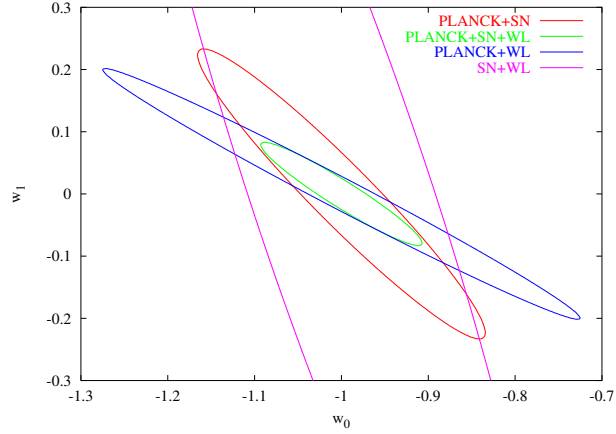}
  \caption{
The $1\sigma$ confidence two-dimensional regions ($\Delta \chi^2=1$)
for the $(w_0,w_1)$ parameters. The plots contrast the
constraints obtained from different combinations of 1 year of data from Planck,
2000 SNe Ia with $z_{max}=0.8$, and a ground-based like WL reference survey with
$f_{sky}=0.7$, $\bar n = 30\, {\rm gal/arcmin}^2$, 
$\left<\gamma_{int}^2\right>^{1/2} \approx 0.4$, and a median
redshift $z_{med}\approx 1$.}
\label{fig:planck}
\end{center}
\end{figure}

The separation of source galaxies into tomographic bins improves
significantly the constraints on cosmological parameters, and
particularly those of dark energy because tomography probes the growth of
structure.  The constraints obtained within different redshift bins
are complementary and add up to reduce the  final uncertainties. We
explore here three tomography studies using two different types
of cosmic shear surveys. For the first survey (ground-based like), the  source
galaxy redshift distribution has a median redshift $z_{median}=1.0$,
and we assume that the photometric redshift knowledge will allow one
to  split the source galaxies into successively 2 and 5 bins. 
For the second survey (space-based like), the source galaxy redshift distribution has a
median redshift $z_{median}=1.5$, and we assume a
good knowledge of photometric redshifts and split the sources into 10
tomographic bins. For each redshift bin $i$, the weighting function is
given by  
\begin{eqnarray}
g_i(\chi) &=&
\left\{
\begin{array}{ll}
{\displaystyle
\int_{\chi_{i}}^{\chi_{i+1}}d\chi' n_i(\chi')\frac{\chi'-\chi}{\chi'}},
& \chi \le \chi_{i+1} \\
0, & \chi > \chi_{i+1}
\end{array}
\right.
\end{eqnarray}
where $n_i(\chi)$ is the bin normalized redshift distribution.  The
average number density of galaxies in this bin is $\Phi_i \bar{n}$
with the fraction $\Phi_i$ given by  
\begin{equation}
\Phi_i=\int_{\chi_{i}}^{\chi_{i+1}} d\chi' n(\chi').
\end{equation}

For example, in the 2 bin case, the normalized distributions  are given by
\begin{eqnarray}
n^{A}(z)&=&
\left\{
\begin{array}{ll}
{\displaystyle
 \frac{n(z)}{1-5/e^{2}}} &
   {z_{p} \le 2 z_{0}}, \\
{0, } & 
  {z_p > 2 z_0},
\end{array}
\right.
\label{eq:z_distA}
\end{eqnarray}
and
\begin{eqnarray}
n^{B}(z)&=&
\left\{
\begin{array}{ll}
{\displaystyle
0,}  & 
  {z_p \le 2 z_0}, \\
{\frac{n(z)}{5/e^{2}}} &
  {z_{p} > 2 z_{0}}.
\end{array}
\right.
\label{eq:z_distB}
\end{eqnarray}
For the 5 bins, we use redshift intervals of $\Delta z=0.6$
over the redshift range $0.0 < z < 3.0$, and for the 10 bins we 
use intervals of $\Delta z=0.3$. 
The convergence power spectra for the 10 bins
with the respective sample variance errors averaged
over bands in $\ell$ are shown in  figure
\ref{fig:convergence}. 
\subsection{Fisher matrices for weak lensing}
\begin{figure}
\begin{center}
\includegraphics[width=2.3in,angle=-90]{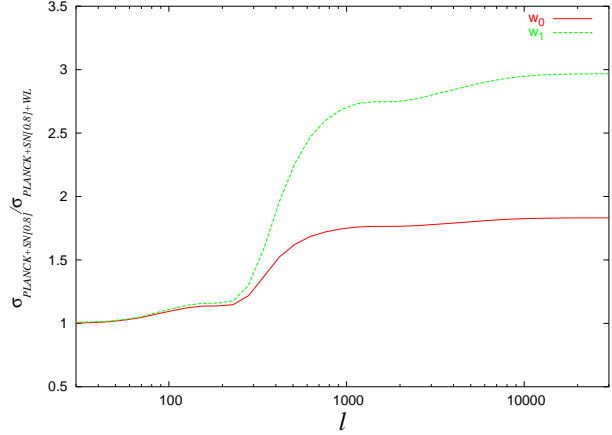}
\caption{
The improvement of the constraints as a function of the scales probed
by the ground-based like weak lensing survey (no-tomography). The
combination PLANCK1+SN[0.8]+WL is used here. The curves represent
$\sigma_{_{PLANCK1+SN[0.8]}}/\sigma_{_{PLANCK1+SN[0.8]+WL}}$ for $w_0$
and $w_1$. The improvement from WL arises from probing
non-linear scales ($l\gtrsim 100$), with a significant jump in the improvement 
between $l\sim 200$ and $\sim 2000$ for the reference survey
considered. }
\label{fig:nonlinear}
\end{center}
\end{figure}
If the convergence field is Gaussian, and the noise is a combination
of Gaussian shape and instrument noise with no intrinsic correlations,
the Fisher matrix is given by:
\begin{equation}
F_{\alpha \beta} = {\sum_{\ell=\ell_{\rm min}}^{\ell_{\rm
        max}}{\frac{1}{(\Delta P_\kappa
        )^2} {\frac{\partial P_\kappa}{\partial p^{\alpha}}}}
        {\frac{\partial P_\kappa}{\partial p^{\beta}}}};
\label{eq:fisher1}
\end{equation}
where the uncertainty in the observed weak lensing spectrum is
given by: \cite{1992ApJ...388..272K,1998ApJ...498...26K}

\begin{equation}
\Delta P_{\kappa}(\ell)=
\sqrt{\frac{2}{(2\ell +1)f_{sky}}}\left (
P_{\kappa}(\ell) + {\left< \gamma_{int}{}^2 \right>\over {\bar n}} \right ) \,,
\label{delta_kappa}
\end{equation}

\noindent where $f_{sky} = \Theta^2 \pi/129600 $ is the fraction of the
sky covered by a survey of dimension $\Theta$ in degrees,  and
$\left<\gamma_{int}^2\right>^{1/2}$ is the intrinsic ellipticity of
galaxies. Guided by previous studies and taking into consideration the
major difficulty of constraining both $w_0$ and $w_1$, we considered
an almost full sky ground-based like survey with $f_{sky}=0.7$, a median
redshift of roughly 1, an average galaxy number density of $\bar n =
30\; {\rm gal/arcmin}^2$, and $\left<\gamma_{int}^2\right>^{1/2}=0.4$.
We also model an ambitious space-based like survey with
$f_{sky}=0.01$, $0.10$, and $0.70$, a median redshift of roughly 1.5, $\bar n = 100\;{\rm
  gal/arcmin}^2$, and $\left<\gamma_{int}^2\right>^{1/2} \approx
0.25$. We have used $\ell_{\rm max}=3000$ since on smaller scales, the
assumption of a Gaussian shear field underlying Eq. (\ref{eq:fisher1})
and the {\sc halofit} approximation to the nonlinear power spectrum
may not be valid for larger $\ell's$. For the minimum $\ell$, we take the fundamental mode
approximation:
\begin{equation}
\ell_{\rm min} \approx \frac{360\rm ~deg}{\Theta} = \sqrt{\pi\over f_{sky} },
\label{eq:lmin}
\end{equation}
i.e. we consider only lensing modes for which at least one wavelength
can fit inside the survey area. For tomography, the Fisher matrix is
generalized using
\begin{equation}
 F_{\alpha\beta} = \sum_{\ell_{\rm min}}^{\ell_{\rm max}}
                (\ell + 1/2) f_{\rm sky}
 {\rm Tr}\left( {\bf C}_\ell^{-1} {\partial{\bf P}_\ell\over\partial p^\alpha}
                       {\bf C}_\ell^{-1} {\partial{\bf P}_\ell\over\partial p^\beta} \right),
\label{eq:fisher2}
\end{equation}
where ${\bf C}_\ell$ is the covariance matrix of the multipole moments
of the observables ${C}^{\kappa \kappa'}_\ell = P_\ell^{\kappa \kappa'}
+ N_\ell^{\kappa \kappa'}$ with $N_\ell^{\kappa \kappa'}=\delta_{\kappa \kappa'}
{\left< \gamma_{\rm int}^2\right > / {\Phi_i\bar{n} }}$ the power spectrum
of the noise in the measurement. 
\begin{figure}
\begin{center}
\includegraphics[width=2.3in,angle=-90]{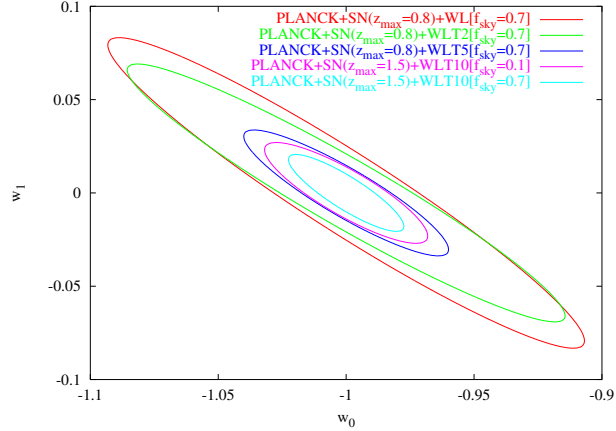}
  \caption{
Tomography and the equation of state parameters: the $1\sigma$ confidence two-dimensional regions ($\Delta \chi^2=1   $)
for the parameters $(w_0,w_1)$. The plots contrast
the constraints from different combinations of 1 year of data from
Planck, 2000 uniformly distributed supernovae with
$z_{max}=0.8,\,1.5$, the ground-based like lensing survey (WL), the
ground-based like survey with successively 2 bin tomography (WLT2) and 5 bin tomography (WLT5), and the deep
space-based like survey with 10 bin tomography (WLT10).} 
\label{fig:w_tomography}
\end{center}
\end{figure}

\subsection{Weak lensing systematic effects}\label{sec:systematics}
The probes considered here have systematic errors and nuisance factors
that need to be well understood and well controlled in order for these
constraints to be achievable. For cosmic shear, several systematic
effects have been identified so far,  see \cite{RefregierARAA} 
and references therein for an overview.

In this analysis, we included the effect of the shear calibration 
bias \cite{Erben,Bacon}\\
\cite{2003MNRAS.343..459H,2002AJ....123..583B,2000ApJ...537..555K,Van} 
on our results by marginalizing over its parameters. In this bias, the
shear is systematically over or under-estimated by a multiplicative
factor, and mimics an overall rescaling of the shear power
spectrum. We used  the absolute power calibration parameter $\cals$
and the relative calibration parameter $\calr$ between two redshift
bins, following the parametrization used and discussed in
\cite{Ishak2004}. The calibration bias  is not detected by the usual
systematic tests such as the E and B modes decomposition of the shear
field and the cross-correlation of the shear maps against the
point-spread function (PSF) maps. In a weak lensing survey, $\cals$
and $\calr$ are parameters of the experiment that can in principle be
determined by detailed simulations of the observations. We impose in
this analysis a reasonable Gaussian prior of 0.04 on these parameters.   

Another serious systematic is the incomplete knowledge of the source
redshift distribution \cite{2000Natur.405..143W} \\
\cite{IshakHirata} including
the redshift bias and scattering. One can argue 
that as long as the scatter in the redshift ($\sigma(z)\approx 0.05$) is
much smaller than the width of the redshift bin $\Delta z=0.3$ (for our 10
bin tomography) the effect on the integrated  results should be
small. This is due to the fact that the scatter will only change
slightly the shape of the edges of the window function that is used
for the redshift integration within each bin. At the contrary, the
redshift bias alters the overall distribution and has been known to
affect significantly the cosmological parameter estimation
\cite{RefregierARAA,Ishak2004,Tereno}. A remedy to this poor knowledge
of the redshift distribution using spectroscopic redshift has been
explored recently in \cite{IshakHirata}. In the present analysis we
marginalize over the redshift bias by including the characteristic
redshift of the distribution as a systematic parameter $z_s$ and we
assume a reasonable Gaussian prior of 0.05 on this parameter. 

It is important to question our assumption of gaussianity for 
the two weak lensing systematic errors that we considered here.
So far, studies have made this assumption. 
The motivation for this assumption is the simplicity of the 
aproximation but its justification needs to be addressed in 
dedicated lensing systematic studies. Thus our treatment of 
these two systematics is only valid under this assumption of 
gaussianity.
For the redshift bias, one has to stress the requirement for 
a sufficiently large number of spectroscopic redshifts, see e.g. \cite{IshakHirata}, 
and a sufficiently large number of source galaxies in order to reduce the 
uncertainties to a point where they can be approximated by 
Gaussians. Also, more narrow band colors and more accurate 
magnitudes (i.e. deeper exposure) are necessary in order to 
break degeneracies between the photometric redshifts.
For the shear calibration bias, the errors cannot be made small
enough by adding more data, and in this case simulations are necessary 
in order to estimate and address the non-gaussianity question.
Massive sky image simulations on which 
the shear is measured and compared to the input will be necessary and  
such studies are planned to be carried out within ongoing 
projects in the lensing communauty, such as 
the Shear TEsting Programme (STEP) project \cite{step}. 
Indeed, studying the shear calibration bias using massive simulations 
is one of the goals of this project \cite{step}.
Hence, the reader should be aware that our treatment of shear 
calibration errors is valid only under the gaussianity assumption,  
and this needs to be checked by future simulation studies.   

Finally, it is important to recall that there are systematic effects 
that were not included in this analysis (e.g. intrinsic alignments
of galaxies,  
uncertainties associated with the non-linear mapping of the matter power
 spectrum, see \cite{RefregierARAA}
and references therein for a list of other effects), however it is encouraging 
to note that many efforts are spent  in
order to study these and other lensing systematic effects and some
progress has been made. With a better understanding of these limiting
factors it will be possible to parameterize them and evaluate their
effect on the cosmological parameter estimation. 
\section{Probing dark energy with supernovae type Ia}\label{sec:snIa}
Supernovae of type Ia are powerful probes of dark energy, as when
properly calibrated they become cosmological standard
candles that can be used to measure distances as a function of
redshift. The luminosity distance to a SN Ia is given by
\begin{equation}
d_{L}\equiv \sqrt{\frac{L}{4 \pi F}}
\end{equation}
where $L$ is the intrinsic luminosity and $F$ is the observed flux.
The apparent magnitude of this SNa Ia can be written as
\begin{equation}
m = 5\: \log_{10}(D_L) + \mathcal{M}
\label{eqn:app_mag}
\end{equation}
where $D_L\equiv H_0 d_L/c $ is the dimensionless luminosity
distance, $\mathcal{M}\equiv M-5\: \log_{10}(H_0/c)+constant$ is
the magnitude parameter, and $M$ is the absolute magnitude,
degenerate here with the Hubble parameter.
In a spatially flat model
\begin{equation}
D_L(z) = (1+z) \; \int_0^z \frac{1}{\sqrt{(1-\Omega_\Lambda)(1+z')^3 +
    \Omega_\Lambda\mathcal{E}(z')}} dz',
\label{eqn:DL_vs_z}
\end{equation}
where  $\mathcal{E}(z)$ is as defined in Eq.(\ref{eq:E_z}).
We use the Fisher matrix for the SNe Ia defined as (see, e.g. \cite{snfisher,HutererTurner})
\begin{equation}
F_{\alpha \beta} = {\sum_{i=1}^{N}}
        {\frac{1}{\sigma_m(D_{L,i})^2}
        {\frac{\partial D_{L,i} }{\partial p^{\alpha}}}}
        {\frac{\partial D_{L,i}}{\partial p^{\beta}}}.
\label{eq:snfisher}
\end{equation}
We use two sets of 2000 SNe Ia uniformly distributed with
$z_{max}=0.8$ and $z_{max}=1.5$.  It is important to briefly note here
that there are systematic uncertainties associated with supernova
searches: these include  luminosity evolution, gravitational lensing
and dust; see, e.g. \cite{SNAP2003b} and references therein. In order
to partly include the effect of these systematics and the effect of
the supernova peculiar velocity uncertainty \cite{Tonry}, we follow
\cite{Kim2003,upadhye} and use the following expression  for the effective
magnitude uncertainty 
\begin{equation}
\sigma_m^{eff}=\sqrt{\sigma_m^2+\left(\frac{5\sigma_v}{ln(10)cz}\right)^2+N_{\{per\;bin\}}\delta_m^2}
\label{eq:snquadrature}
\end{equation}
where, a peculiar velocity of $\sigma_v=500km/sec$ is assumed, and
following \cite{Kim2003,upadhye} we use $\delta_m=0.02$ for
space-based supernova survey data, and assume $\delta_m=0.04$ for
ground-based survey data. The quadrature relation 
(\ref{eq:snquadrature}) assures that there is an uncertainty floor 
set by the systematic limit $\delta_m$ so that the overall uncertainty 
per bin cannot be reduced to arbitrarily low values by adding more supernovae.
\section{Probing dark energy with CMB and cosmic complementarity}\label{sec:cmb}
\begin{table*}
  \begin{center}
  \caption{Summary of the parameter $(w_0,w_1)$ estimation errors
    ($1\sigma$ uncertainties) 
    from different combinations of probes. CMB experiments are WMAP 8
    years, Planck 1 year, and ACT 1 year (unlensed spectra) combined with WMAP 8
    years. 
(We did not include 
 constraints from the Sunayev-Zeldovich effect or lensing of
the CMB that ACT and PLANCK will be useful with.) 
The 2000 supernovae are uniformly distributed with
    $z_{max}=0.8$. WL is for a ground-based like weak lensing survey with $f_{sky}=0.7$,
    $\bar n = 30\, {\rm gal/arcmin}^2$, 
    $\left<\gamma_{int}^2\right>^{1/2} \approx 0.4$, and a median
redshift $z_{med}\approx 1$. WLT2 refers to
    the same weak lensing survey but with 2 bin tomography.
    The best constraints from the combinations in this table are from PLANCK1+SN[0.8]+WLT2.
}
  \label{tab:w_1}
%\begin{ruledtabular}
  \begin{tabular}{ccccccccccc}
\hline
             &CMB   &+SN  &+WL &+WLT2&+SN+WL&+SN+WLT2&SN+WL &SN+WLT2\\
             &         &     &    &     &      &        &no-CMB&no-CMB \\\hline%\\%[3pt]
             &WMAP-8 alone&     &    &     &      &        &      &       \\
$\sigma(w_0)$& 3.73    &0.25 &0.66& 0.35&  0.21& 0.11   & 0.24 &  0.11 \\
$\sigma(w_1)$& 5.65    &0.87 &1.45& 0.66&  0.59& 0.26   & 0.93 &  0.35 \\\hline
             &WMAP8+ACT&     &    &     &      &        &      &       \\
$\sigma(w_0)$& 0.82    &0.20 &0.58& 0.22& 0.18 & 0.11   &  "   &  "    \\
$\sigma(w_1)$& 1.87    &0.59 &1.40& 0.42& 0.48 & 0.25  &   "  &   "   \\
\hline
             &PLANCK-1 alone&     &    &     &      &        &      &       \\
$\sigma(w_0)$& 0.50    &0.17 &0.28& 0.23& 0.093& 0.086  &  "   &  "    \\
$\sigma(w_1)$& 0.31    &0.23 &0.20& 0.18& 0.083& 0.069  &  "   &  "    \\\hline
 \end{tabular}
%\end{ruledtabular}
 \end{center}
\end{table*}
CMB is a powerful cosmological probe; however, like other probes it
suffers from some parameter degeneracies and needs to be combined with
other data sets in order to provide tight constraints on  dark energy
parameters. For example, it is well known that cosmic shear and CMB
have different types of degeneracies in their parameters, which are
nicely broken when these probes are combined. Indeed, among the
orthogonal directions of degeneracy between cosmic shear and CMB, are
the known doublets $(\Omega_m,\sigma_8)$, $(h,n_s)$, and
$(n_s,\alpha_s)$; see, e.g. \cite{Tereno}. We use this cosmic 
complementarity in the present analysis where the statistical error
(with some systematics included) on a given parameter $p^{\alpha}$  is
given by: 
\begin{equation}
\sigma^2(p^{\alpha})\approx [({\bf F}_{CMB}+{\bf
F}_{WL}+{\bf F}_{SNe}+\Pi)^{-1}]^{\alpha \alpha},
\label{eq:sigma}
\end{equation}
where $\Pi$ is the prior curvature matrix, and ${\bf F}_{CMB}$, ${\bf
  F}_{WL}$ and ${\bf F}_{SNe}$ are the Fisher matrices from CMB, weak
  lensing, and supernovae, respectively. We only impose priors on the
  characteristic redshift and the calibration parameters by taking
  priors of $\sigma(z_p)=0.05$ and $\sigma(\cals)=\sigma(\calr)=0.04$
  on the calibration parameters (corresponding to 2\% rms uncertainty
  on the amplitude calibration; \cite{2003MNRAS.343..459H}). We may
  add the CMB and weak lensing (WL) Fisher matrices together because
  the primary CMB anisotropies are generated at much larger comoving
  distance than the density fluctuations that give rise to weak
  lensing, hence it is a good approximation to take them to be
  independent. For CMB, we project constraints from 8 years of WMAP
  data (WMAP-8), 1 year of PLANCK data (PLANCK-1), and 1 year of ACT
  data  combined with WMAP-8. The experiment specifications used for
  Planck and ACT are listed in table (\ref{tab:cmbexp}). 
For 8 year WMAP, we include $TT$, $TE$, and $EE$ power spectra,
  assuming $f_{sky}=0.768$ (the Kp0 mask of Ref.~\cite{Bennet2}),
  temperature noise of $400$, $480$, and $580\,\mu$K~arcmin in Q, V,
  and W bands respectively (the rms noise was multiplied by $\sqrt{2}$
  for polarization), and the beam transfer functions of
  Ref.~\cite{Page}.
\section{Results and discussion}
We calculated future constraints on dynamical dark energy parameters
obtained form several combinations of cosmic microwave background
experiments (CMB), supernova searches (SNe Ia), and weak lensing
surveys (WL) with and without tomography. For CMB, we considered
future constraints from 8 years of data from WMAP (WMAP8),  one year
of data from Planck (PLANCK1),  and one year of data from the Atacama
Cosmology Telescope (ACT1). We used two sets of 2000 supernovae with
$z_{max}=0.8$ (SN[0.8]) and $z_{max}=1.5$ (SN[1.5]) respectively, and
considered various cosmic shear reference surveys: an almost full sky
(70\%) ground-based like survey (WL) with successively 2 (WLT2) and 5 (WLT5)
tomographic bins; a deep space-based like survey with 10 
tomographic bins (WLT10) covering  successively 1\%, 10\% and 70\% 
of the sky. 
We combined these experiments in doublets and triplets,
taking into account space-based or ground-based like experiments for
supernovae and weak lensing. One should take the uncertainties
obtained on the dark energy parameters from  CMB-only Fisher matrices
with some caution as some of them are large and the Fisher matrix
approximation may not be valid. We compared our CMB-only constraints
for Planck with  those of reference \cite{Hu01} and found them in good
agreement. For example, when we fix the parameters $w_1$ to compare
with reference \cite{Hu01}, we find that our
$\{\sigma(\Omega_{\Lambda})=0.087,\sigma(w)=0.31\}$ are in good agreement
with $\{\sigma(\Omega_{\Lambda})=0.098,\sigma(w)=0.32\}$ from
\cite{Hu01}. Importantly, the constraints we obtain from any
combination of the three probes are significantly smaller then
CMB-only and therefore the constraints obtained are good estimates of
the low bound of the uncertainties. 
Our results are summarized in tables \ref{tab:w_1}, \ref{tab:w_a}, and
\ref{tab:compare} and in figures \ref{fig:wmap}, \ref{fig:planck}, \ref{fig:nonlinear}, 
and \ref{fig:w_tomography}. We looked
for combinations of experiments that will  provide constraints that
are small enough to  answer conclusively some of the dark energy
questions. 

\begin{table*}
  \begin{center}
  \caption{Same as table \ref{tab:w_1} but for the dark energy parameterization $(w_0,w_a).$
As usual
 the errors on $w_a$ are larger than those on $w_1$
(roughly equal to twice the errors on $w_1$).  The best constraints in this table are from PLANCK1+SN[0.8]+WLT2.}
  \label{tab:w_a}
%\begin{ruledtabular}
  \begin{tabular}{ccccccccccc}
\hline
             &CMB      &+SN  &+WL &+WLT2&+SN+WL&+SN+WLT2&SN+WL &SN+WLT2\\
             &         &     &    &     &      &        &no-CMB&no-CMB \\\hline%\\%[3pt]
             &WMAP-8 alone&     &    &     &      &        &      &       \\
$\sigma(w_0)$& 1.84    &0.24 &0.46&0.39 &  0.21& 0.14   & 0.26 & 0.14  \\
$\sigma(w_a)$& 3.03    &1.25 &1.61&1.24 &  0.92& 0.53   & 1.37 & 0.76  \\\hline
             &ACT+WMAP8&     &    &     &      &        &      &       \\
$\sigma(w_0)$& 0.52    &0.21 &0.27&0.42 & 0.19 & 0.14   &  "   &   "   \\
$\sigma(w_a)$& 1.83    &0.92 &0.85&1.48 & 0.76 & 0.50   &  "   &   "   \\
\hline
             &PLANCK-1 alone&     &    &     &      &        &      &       \\
$\sigma(w_0)$& 0.67    &0.15 &0.28& 0.24& 0.097& 0.088  &  "   &   "   \\
$\sigma(w_a)$& 0.52    &0.30 &0.32& 0.29& 0.133& 0.111  &  "   &   "   \\\hline
  \end{tabular}
%\end{ruledtabular}
 \end{center}
\end{table*}

The first question that one would like to answer is whether dark energy is
a cosmological constant or a dynamical component. 
The most decisive answer will be to rule out significantly the cosmological constant.
A less decisive but very suggestive answer will be to show 
that dark energy parameters are those of a cosmological 
constant with a few percent uncertainty only.
This can be compared to the case of spatial curvature in
the universe: when CMB results constrained the total energy density
to $\Omega_T=1.02\pm 0.02$ at the one sigma level, then it became
generally more accepted that  spatial curvature is negligible. 
However, it is important to note that it will remain always possible 
to build dark energy models that could have a set of parameters 
indistinguishable from those of a cosmological constant within the limits set.
So in this scenario other types of tests than the
equation of state will be required in order to close the question. 

As shown in table \ref{tab:compare}, the combinations
PLANCK1+SN[1.5]+WLT10[$f_{sky}=0.1$] and PLANCK1+SN[0.8]+WLT5[$f_{sky}=0.7$] 
provide impressive constraints that reach the goal set.
This is followed by the constraints from PLANCK1+SN[1.5]+ WLT10[$f_{sky}=0.01$].
One should note here that for the equation of state parameters, only small additional improvements to these constraints
are obtained when an extremely ambitious sky coverage of 70\% 
is considered for PLANCK1+SN[1.5]+WLT10 [$f_{sky}=0.7$].
Finally, the constraints on the equation of state parameters from
PLANCK1+SN[0.8]+WLT2[$f_{sky}=0.7$] are 
 not small enough for the criterion of a few percent set above.

Another test to answer the same question above is by probing directly
the dark energy density at various redshift points. For example, if
future data will show significant 
departures of the parameters $\mathcal{E}_1$ or $\mathcal{E}_2$ (see
Sec. \ref{sec:de_param}), from unity then a cosmological constant can
be ruled out conclusively. Our analysis shows that this might be a
less difficult test  as even PLANCK+SN[0.8]+WLT5[$f_{sky}=0.7$] has the 
potential to achieve  $\sigma(\mathcal{E}_1)=0.012$ and
$\sigma(\mathcal{E}_2)=0.049$.

A second question of interest is what combination of experiments could
distinguish between some currently proposed models of dark energy. Of
particular interest are models that predict an equation of state with
parameters that are significantly different from those of a
cosmological constant. For example, one could consider quintessence
tracker models \cite{zlatev} or super-gravity inspired models
\cite{brax}, for which, $w_0 \gtrsim -0.8$ 
 and $dw/dz(z=0)\sim 0.3$. From our
tables, we see that the distinction between these models and a
cosmological constant can be achieved by several different
combinations of experiments with different levels of precision.

Our tables \ref{tab:w_1} and \ref{tab:w_a} show that after combining
Planck and supernova constraints, weak lensing without tomography adds
an improvement of roughly a factor of 2 or better to the
constraints. As shown in figure \ref{fig:nonlinear}, the WL-improvement
arises from probing non-linear scales ($l\gtrsim 100$), with a
significant jump between $l\sim 200$ and $\sim 2000$ for the
ground-based like survey considered. Adding 2 bins tomography
to the lensing survey  provides an additional factor of 2
improvement to the combination WMAP8+SN[0.8]+WL and to
combination ACT1+WMAP8+SN[0.8]+WL.
We mention here that these results do not
include constraints from Sunayev-Zeldovich effect or lensing of
the CMB, that ACT and PLANCK will be useful with.

In table \ref{tab:compare} and figure \ref{fig:w_tomography}
 we summarize our results on multiple-bin tomography.
 The constraints on the equation of state
parameters from PLANCK1+SN[1.5] improve by factors 3-5 when WLT10 with $f_{sky}=0.1$ is
added to the combination. Also, the constraints on the equation of state
parameters from PLANCK1+SN[0.8]+WLT5[$f_{sky}=0.7$] are roughly factors of 3-6
better than those obtained from PLANCK1+SN[0.8]. 
Thus, we find that the improvements obtained from multiple-bin tomography
lensing surveys are important for the questions raised at the beginning
of this section as it brings the uncertainties significantly closer to
the goal of a few percent.
Therefore, the present study shows that tomography
is very useful to add further improvements to the constraints on dark
energy parameters using both 
 ground-based experiments and space-based experiments. The precise 
discussion of the technical and instrumental feasibility of multiple-bin 
tomography from ground or from space is beyond the scope of this 
paper and should be addressed somewhere else.

We took care to include some systematic effects in our analysis. We
parameterized the weak lensing calibration bias and assumed reasonable
priors of 0.04 on the calibration parameters. We also parameterized the
redshift bias and used a reasonable prior of 0.05. However, there are
other systematic effects that we did not include and that may affect
our results. For example, we did not include the effect of intrinsic
correlations between the lensing source galaxies on our results 
\cite{2000ApJ...545..561C,2000MNRAS.319..649H,2000ApJ...532L...5L,2001MNRAS.320L...7C,2001ApJ...559..552C,2002MNRAS.333..501B}\\
\cite{2002MNRAS.335L..89J,2003AJ....125.1014J,2004MNRAS.347..895H,HirataSeljak2004},
 and we used the {\sc{HALOFIT}}  fitting formula to evaluate the
 non-linear matter power spectrum (full simulations should be used for
 real data analysis). 
As discussed in section 3.4, our inclusion of these two lensing 
systematics assumed their gaussianity. The effect of these systematics 
on our results is thus valid only under this approximation. 
Future dedicated studies of weak lensing systematic effects 
should address the issue of non-gaussianity. 
Also, following previous work, we included in
 our supernova constraints a conservative systematic limit, but more
 studies need to be done there as well. 

Nevertheless, our results show that after the combination of CMB,
supernovae, and weak lensing surveys, tomography with 
very large fractions of the sky and many redshift bins has the potential to
add key improvements to the dark energy parameter constraints by
bringing them to the level of a few percent.

On the other hand, the requirement for very ambitious and sophisticated 
surveys in order to achieve some of these constraints, and the difficulty
to obtain any further significant improvements, even with the most ambitious 
survey we considered, suggest the need for new tests to probe the nature 
of dark energy in addition to constraining its equation of state.

\begin{table*}
{\scriptsize
  \begin{center}
  \caption{Various tomography analyses: a comparative summary of the constraints 
  ($1\sigma$ uncertainties) on the different dark energy parameterizations from
    Planck-1, 2000 uniformly distributed supernovae with
    $z_{max}=0.8,1.5$, the ground-based like lensing survey with successively 2 bin
    tomography (WLT2) and 5 bins (WLT5), and the deep space-based like survey with 10 bin
    tomography (WLT10). The results are presented for the dark energy
    parameters $\{w_0, w_1\}$, $\{w_0, w_a\}$ and
    $\{\mathcal{E}_1\equiv \rho_{de}(z=0.5)/\rho_{de}(0),
    \mathcal{E}_2\equiv \rho_{de}(z=1.0)/\rho_{de}(0)\}$ 
}
  \label{tab:compare}
%\begin{ruledtabular}
  \begin{tabular}{cccccccc}
\hline
&PLANCK-1&+SN($z_{max}=1.5$)&+SN($z_{max}=0.8$)&+SN(0.8)+WLT5&+SN(1.5)+WLT10&+SN(1.5)+WLT10&+SN(1.5)+WLT10\\

&alone & &+WLT2[$f_{sky}=0.7$]&[$f_{sky}=0.7$]&[$f_{sky}=0.01$]&[$f_{sky}=0.1$]&[$f_{sky}=0.7$]\\\hline

$\sigma(w_0)$& 0.50  &0.11  &0.086 &0.04 &0.048 &0.032 & 0.023\\
$\sigma(w_1)$& 0.31  &0.13  &0.069 &0.034&0.042 &0.027 & 0.021\\\hline

$\sigma(w_0)$& 0.67  &0.12  &0.088 &0.041&0.049 &0.033 & 0.023\\
$\sigma(w_a)$& 0.52  &0.21  &0.111 &0.056&0.067 &0.040 & 0.026\\\hline

$\sigma(\mathcal{E}_1)$&0.11&0.048&0.029&0.012 &0.013 &0.010& 0.009\\
$\sigma(\mathcal{E}_2)$&0.32&0.12 &0.065&0.049 &0.082 &0.040& 0.018\\
\hline
  \end{tabular}
%\end{ruledtabular}
 \end{center}
}
\end{table*}

\section*{acknowledgments}
The author thanks David Spergel for support and for useful discussions.
The author thanks N. Bahcall, O. Dor\'e, C. Hirata, P. Steinhardt,  A. Upadhye, and L. Van Waerbeke for
useful comments. The author acknowledges the support of the Natural
Sciences and Engineering Research Council of Canada (NSERC) and the 
support of NASA Theory Award NNG04GK55G.

\end{document}